\documentclass[12pt]{article} 
\usepackage[sectionbib]{natbib}
\usepackage{array,epsfig,fancyheadings,rotating}
\usepackage[]{hyperref}  
\usepackage{sectsty, secdot}
\sectionfont{\fontsize{12}{14pt plus.8pt minus .6pt}\selectfont}
\renewcommand{\theequation}{\thesection\arabic{equation}}
\subsectionfont{\fontsize{12}{14pt plus.8pt minus .6pt}\selectfont}

\usepackage[margin=1in]{geometry}
\usepackage{amsmath}
\usepackage{amssymb}
\usepackage{amsfonts}
\usepackage{multirow}
\usepackage{amsthm}
\usepackage{mathrsfs}
\usepackage{bm}
\usepackage{graphicx}
\usepackage{tikz}
\usetikzlibrary{positioning,arrows.meta}
\usepackage{subcaption}
\usepackage{caption}
\usetikzlibrary{graphs,positioning,arrows}
\usepackage{float}
\usepackage{empheq}

\newtheorem{theorem}{Theorem}

\newtheorem{assumption}{Assumption}
\newtheorem{mechanism}{Mechanism}
\theoremstyle{definition}

\newtheorem{remark}{Remark}
\begin{document}

\renewcommand{\baselinestretch}{2}

\markright{ \hbox{\footnotesize\rm Statistica Sinica
}\hfill\\[-13pt]
\hbox{\footnotesize\rm
}\hfill }

\markboth{\hfill{\footnotesize\rm FIRSTNAME1 LASTNAME1 AND FIRSTNAME2 LASTNAME2} \hfill}
{\hfill {\footnotesize\rm FILL IN A SHORT RUNNING TITLE} \hfill}

\renewcommand{\thefootnote}{}
$\ $\par

\fontsize{12}{14pt plus.8pt minus .6pt}\selectfont \vspace{0.8pc}
\centerline{\large\bf Mediation Analysis with Multiple Mediators}
\vspace{2pt} 
\centerline{\large\bf Subject to Missing Not at Random}
\vspace{.4cm} 
\centerline{Yanfei Jin, Shanshan Luo, Xueli Wang$^{*}$, Zhi Geng} 
\vspace{.4cm} 
\centerline{\it School of Mathematics and Statistics, Beijing Technology and Business University}
 \vspace{.55cm} \fontsize{9}{11.5pt plus.8pt minus.6pt}\selectfont

\begin{quotation}
\noindent {\it Abstract:}

Causal mediation analysis serves as a key tool for uncovering the mediating mechanisms linking treatments to outcomes. Existing methods for mediation analysis with multiple mediators typically 
assume complete observations or missing-at-random and may yield biased estimation 
when mediator values are missing not at random (MNAR). This paper studies the identification 
and estimation of causal mediation effects with multiple mediators subject to MNAR missingness. 
We consider a broad class of MNAR mechanisms in which missingness may depend on unobserved 
mediators, treatment, covariates, and outcomes. Under a series of increasingly general MNAR mechanisms, we establish identified natural direct and indirect effects, effectively generalizing existing mediation analysis to handle nonignorable missing mediators. Based on the proposed identification framework, we develop estimation 
procedures for causal mediation effects and evaluate their finite-sample performance through 
simulation studies. The results demonstrate satisfactory performance across a range of 
missingness scenarios. An application to data from the National Health and Nutrition Examination 
Survey(NHANES) illustrates the practical utility of the proposed methodology for investigating 
mediation pathways in the presence of nonignorable missing data.

\vspace{9pt}
\par
\end{quotation}\par

\def\thefigure{\arabic{figure}}
\def\thetable{\arabic{table}}

\renewcommand{\theequation}{\thesection.\arabic{equation}}

\fontsize{12}{14pt plus.8pt minus .6pt}\selectfont

\section{Introduction}

In fields such as medicine, social sciences, and economics, researchers have long been interested 
in the causal effects of interventions or exposures on outcomes. However, reporting only the 
overall average effect often fails to reveal the mechanisms through which the intervention 
operates, leaving the underlying causal pathways unclear. Causal mediation analysis has 
therefore emerged as an important tool for characterizing such mechanisms, with the primary goal 
of decomposing the total effect into an indirect effect transmitted through mediators and a 
direct effect not mediated by them.

Under the counterfactual framework, the theoretical foundations of mediation analysis are well 
established. \citet{Pearl2001} first introduced the definitions of natural direct and indirect 
effects within nonparametric structural causal models. 
 \citet{Imai2010a} proposed a unified framework 
for causal mediation analysis that integrates definitions, identification, estimation, and 
sensitivity analysis. \citet{Imai2010b} further established nonparametric identification results 
for the average causal mediation effect under the sequential ignorability assumption. In 
addition, cross-world independence, as a key structural condition for identifying natural direct 
and indirect effects, together with sequential ignorability, forms the theoretical basis of most 
existing mediation analysis methods. 
Including exposure--mediator interactions, semiparametric efficiency theory, and estimation 
approaches based on weighting or imputation \citep{Valeri2013, Tchetgen2012, Huber2014, 
Vansteelandt2012b}.

In the presence of multiple mediators, \citet{Daniel2015} discussed the definition 
and identification of path-specific effects, noting that different decompositions rely on 
distinct counterfactual assumptions and often require strong conditions. \citet{VanderWeele2014} 
pointed out that analyzing each mediator separately may lead to bias when mediators are 
correlated. \citet{Taguri2018} 
proposed a framework for non-ordered mediators, while \citet{Xia2022} developed decomposition 
formulas and multiply robust estimation methods. Furthermore, \citet{Zhou2022} and 
\citet{Chen2025} investigated identification and generalizations of path-specific effects in the 
presence of ordered mediators. However, these studies typically assume that mediators are fully 
observed or at least satisfy missing at random (MAR), where the missingness mechanism is 
independent of unobserved data. They do not systematically address the more challenging missing 
not at random (MNAR) setting, where missingness depends on unobserved mediator values.

In practice, missing data are pervasive in mediation analysis. \citet{Preacher2004} noted that 
missingness in mediators and outcomes is extremely common in psychology and social science 
research. \citet{Rubin1976} classified missing data mechanisms into missing completely at random 
(MCAR), MAR, and MNAR, and showed that under MAR, the missingness mechanism can be ignored for 
inference under certain conditions. Based on this assumption, a rich body of methods has been 
developed, including likelihood-based approaches \citep{Dempster1977, Little2019, Moreno2020}, 
imputation methods \citep{Rubin2004, Kim2011, vanBuuren2018, Sengewald2024}, inverse probability 
weighting \citep{Robins1994, Seaman2013, Sun2018b}, and semiparametric inference 
\citep{Robins1995, Tsiatis2006, Davidian2022}. However, when missingness depends on unobserved 
values (i.e., MNAR), these methods may yield biased results.

Under MNAR, target parameters are generally not point identifiable without additional structural 
assumptions. Existing approaches introduce identifiable structures that can be broadly 
categorized into three classes. First, identification strategies based on external information, 
such as instrumental variable methods \citep{Yang2014a, Wang2014} and shadow variable approaches 
\citep{Sun2018a, Miao2016b}. Second, parametric methods based on distributional assumptions, 
including normal mixture models \citep{Miao2016a} and self-censoring models \citep{Li2023}. 
Third, graphical model approaches that encode structural relationships between the missingness 
mechanism and the data-generating process \citep{Ma2003, Mohan2021}. Meanwhile, some studies 
have explored identification from a causal inference perspective in specific settings.  \citet{Ding2014} studied subgroup causal effects with MNAR covariates, and 
\citet{Yang2019} showed that causal effects can still be nonparametrically identified under 
outcome-independent missingness, even when confounders are MNAR.

In contrast, incorporating MNAR mechanisms into causal mediation analysis remains largely 
underexplored. \citet{Li2017} investigated identification and estimation of causal mediation 
effects under various missingness mechanisms in the outcome. \citet{Huber2020} considered sample 
selection and outcome attrition, but their analysis mainly relies on MAR or instrumental variable 
assumptions. More recently, \citet{Zuo2025} studied identification conditions for direct and 
indirect effects when both mediator and outcome exist MNAR. Nevertheless, existing work 
focuses on single mediator or specific missingness structure. The case of multiple 
mediators with simultaneous nonignorable missingness poses substantially greater identification 
challenges.

Recent studies have increasingly focused on more complex missing-data structures in mediation 
analysis. \citet{Shan2026} investigated causal mediation inference when confounders are subject to 
MNAR, proposing a generalized shadow variable framework. They established theoretical results including asymptotic 
normality, local efficiency, and the semiparametric efficiency bound. However, their work primarily 
addresses nonignorable missingness in confounders and does not consider the joint missingness of 
multiple mediators. \citet{Dashti2025} evaluated the performance of multiple imputation methods  when mediators, outcomes, or intermediate confounders are 
involved in the missingness mechanism. However, 
their analysis is based on the MAR assumption. In addition, \citet{Mhasawade2024} proposed a sensitivity analysis framework under MNAR settings. 
However, their focus is on robustness assessment rather than the identification within a 
counterfactual framework.

To illustrate, consider the effect of obesity on the risk of kidney function impairment. obesity may damage renal microvasculature through chronic inflammation and may also accelerate glomerular 
sclerosis by increasing blood pressure. These mediators—such as inflammatory markers and blood pressure—typically lack a clear causal ordering. In large-scale surveys such as the National 
Health and Nutrition Examination Survey (NHANES), these variables are often subject to missingness due to health conditions or compliance issues, and such missingness is typically 
nonignorable (e.g., more severe patients are more likely to miss examinations). In this setting, the joint distribution of mediators cannot be directly recovered from observed data using existing methods, leading to fundamental identification challenges for counterfactual mediation decomposition of effect.

Motivated by these challenges, this paper studies the identifiability of causal mediation effects 
under MNAR in the presence of multiple  mediators . 
Under a set of interpretable conditional independence structures, we consider four progressively 
more complex MNAR mechanisms. For Mechanisms 1--3, we establish sufficient conditions for 
nonparametric identification of natural direct and indirect effects. For the most general Mechanism 4, we show 
that point identification can still be achieved under appropriate parametric assumptions. The 
remainder of the paper is organized as follows. Section 2 introduces notation and basic 
assumptions. Section 3 describes the missingness mechanisms for mediators. Section 4 presents 
identification results under different MNAR mechanisms. Section 5 evaluates the proposed methods 
via simulation studies. Section 6 applies the methods to National 
Health and Nutrition Examination Survey (NHANES) data. Section 7 concludes with 
a discussion. Proofs and details for the parametric estimation are 
provided in the Supplementary Materials.

\section{Notation and assumptions}

Consider a random sample of size $n$ consisting of independent and identically distributed 
observations from an infinite superpopulation. Let $T$ denote a binary treatment indicator, 
where $T=1$ indicates treatment and $T=0$ indicates control. Let $Y$ be the outcome variable, 
and let $\mathbf{X}$ denote a vector of pre-treatment covariates. We consider two mediators, $M_1$ and $M_2$. For each 
mediator $M_i$ $(i=1,2)$, let $R_i$ denote its missingness indicator, where $R_i=1$ if $M_i$ 
is observed and $R_i=0$ otherwise. $T$ and $R_i$ are binary variables, 
whereas $M_1$, $M_2$, $Y$, and $\mathbf{X}$ may be either discrete or continuous unless 
otherwise specified.

Let $\mathscr{X}$, $\mathscr{M}_1$, $\mathscr{M}_2$, and $\mathscr{Y}$ denote the supports of 
$X$, $M_1$, $M_2$, and $Y$, respectively. We further invoke the standard consistency and 
composition assumptions commonly used in the potential outcomes framework. The consistency 
assumption states that the observed outcome and mediators coincide with their corresponding 
potential values under the observed treatment level, whereas the composition assumption links 
nested counterfactuals to treatment-specific potential outcomes.

For each unit, denote by $M_i(1)$ and $M_i(0)$ the potential mediator values under treatment and 
control, respectively. Under consistency, the observed mediator satisfies 
$M_i=T\cdot M_i(1)+(1-T)\cdot M_i(0)$. Similarly, let $Y(1)$ and $Y(0)$ denote the potential 
outcomes under treatment and control, respectively. More generally, let $Y(t,m_1,m_2)$ denote 
the potential outcome that would be observed if treatment were set to level $t\in\{0,1\}$ and 
the mediators were set to values $m_1\in\mathscr{M}_1$ and $m_2\in\mathscr{M}_2$. The nested 
potential outcome $Y\{1,M_1(0),M_2(0)\}$ represents the outcome that would be observed under 
treatment while both mediators take the values they would attain under control. By the 
composition assumption, the potential outcomes satisfy $Y(1) = Y(1, M_1(1), M_2(1))$ and 
$Y(0) = Y(0, M_1(0), M_2(0))$. 

The natural indirect effect (NIE) through mediator $M_1$ is defined as 
\[
NIE_{M_1} = E\{ Y(1, M_1(1), M_2(0)) - Y(1, M_1(0), M_2(0)) \}.
\]
This quantity represents the average effect of treatment on the outcome operating through 
treatment-induced changes in $M_1$, while holding $M_2$ fixed at its control level. 
Analogously, the NIE through $M_2$ is 
\[
NIE_{M_2} = E\{ Y(1, M_1(0), M_2(1)) - Y(1, M_1(0), M_2(0)) \},
\]
which captures the average effect transmitted through treatment-induced changes in $M_2$, with 
$M_1$ fixed at its control level. The total natural indirect effect, representing the overall 
mediation effect transmitted jointly through $M_1$ and $M_2$, is defined as
\[
NIE = E\{ Y(1, M_1(1), M_2(1)) - Y(1, M_1(0), M_2(0)) \}.
\]

We assume that the potential mediator satisfies $M_1(t,m_2)=M_1(t)$, $M_2(t,m_1)=M_2(t)$
indicating that $M_2$ is not causally affected by $M_1$ and that the two mediators are 
considered parallel. The mediated interactive effect (MI) between $M_1$ and $M_2$ is defined as 
the difference between the total indirect effect and the sum of the mediator-specific indirect 
effects:
\[
\begin{aligned}
MI = \mathbb{E}\bigl\{&
  Y(1, M_1(1), M_2(1)) - Y(1, M_1(1), M_2(0)) \\
&- Y(1, M_1(0), M_2(1)) + Y(1, M_1(0), M_2(0))
\bigr\}.
\end{aligned}
\]
Thus, the total indirect effect can be decomposed additively into the indirect effects through 
$M_1$ and $M_2$ and the mediated interactive effect between $M_1$ and $M_2$, that is 
$NIE = NIE_{M_1} + NIE_{M_2} + MI$. 

The natural direct effect (NDE) is defined as
\[
NDE = E\{ Y(1, M_1(0), M_2(0)) - Y(0, M_1(0), M_2(0)) \}.
\]
The NDE quantifies the effect of treatment on the outcome that is not mediated through either 
$M_1$ or $M_2$, while fixing both mediators at their control values. Consequently, the average 
treatment effect, $ATE=E\{Y(1)-Y(0)\}$, can be decomposed as $ATE=NDE+NIE.$

Let $A \perp\!\!\!\perp B \mid C$ denote conditional independent between the random variables 
$A$ and $B$  given the random variable $C$.

\begin{assumption}[No Treatment–Outcome Confounding]
There is no unmeasured confounding between the treatment and the potential outcome, formally 
given by
\[ 
T \perp\!\!\!\perp Y(t, m_1, m_2) \mid X
\]
for all $t$, $m_1$, and $m_2$.
\end{assumption}

\begin{assumption}[No Mediator–Outcome Confounding]
There is no unmeasured confounding between the mediators and the potential outcome, formally 
given by
\[
(M_1(t), M_2(t)) \perp\!\!\!\perp Y(t, m_1, m_2) \mid (X, T=t)
\]
for all $t$, $m_1$, and $m_2$.
\end{assumption}

\begin{assumption}[No Treatment–Mediator Confounding]
There is no unmeasured confounding between the treatment and the mediators, formally given by
\[
T \perp\!\!\!\perp (M_1(t), M_2(t)) \mid X
\]
for all $t$.
\end{assumption}

\begin{assumption}[Cross-World Independence]
There exists cross-world independence between the potential outcomes and the potential mediators:
\[
Y(t, m_1, m_2) \perp\!\!\!\perp (M_1(t'), M_2(t'')) \mid X,
\]
\[
(M_1(t') \perp\!\!\!\perp M_2(t'')) \mid X
\]
for all $t$, $t'$, $t''$, $m_1$, and $m_2$.
\end{assumption}

\section{Missing data mechanisms}  
To identify the NIE and NDE, we first introduce the mediation formulas under complete data.

\begin{theorem}
Suppose that $M_1$, $M_2$, and $X$ are continuous. Under Assumptions 1--4, 
for $t,t',t''\in\{0,1\}$, the following mediation formula holds:
\[
\begin{aligned}
&\mathbb{E}\left\{ Y(t, M_1(t'), M_2(t'')) \right\}  \\
&= \int_{\mathscr X} \int_{\mathscr M_1} \int_{\mathscr M_2}
\mathbb{E}\left( Y \mid T=t, M_1=m_1, M_2=m_2, X=x \right) \\
&\quad \times f(m_1 \mid T=t', X=x)
       f(m_2 \mid T=t'', X=x)
       f(x)
       \, dm_2 \, dm_1 \, dx .
\end{aligned}
\]
Equivalently, conditional on $X=x$, we have
\[
\begin{aligned}
&\mathbb{E}\left\{ Y(t, M_1(t'), M_2(t'')) \mid X=x \right\} \\
&= \int_{\mathscr M_1} \int_{\mathscr M_2}
\mathbb{E}\left( Y \mid T=t, M_1=m_1, M_2=m_2, X=x \right) \\
&\quad \times f(m_1 \mid T=t', X=x)
       f(m_2 \mid T=t'', X=x)
       \, dm_2 \, dm_1 .
\end{aligned}
\]
\end{theorem}

When $M_1$, $M_2$, and $X$ are discrete, the same identification result holds 
with the integrals replaced by the corresponding summations over the support of 
$(M_1,M_2,X)$, and the density functions replaced by the corresponding 
probability mass functions.

When some mediator values are missing, identification of the NIE and NDE requires recovering 
the conditional distributions $P(Y \mid M_1, M_2, T, X)$, $P(M_1 \mid T, X)$, and 
$P(M_2 \mid T, X)$, or equivalently, the joint distribution $P(Y, M_1, M_2 \mid T, X)$, from 
the incompletely observed data.

In randomized experiments where mediator values $M_i$ may be missing, the missingness mechanism 
of the response indicators $R_i$ plays a crucial role in determining the identifiability of the 
NIE and NDE. When the response indicators are independent of all study variables, that is, 
$(R_1, R_2) \perp\!\!\!\perp (T, M_1, M_2, Y, X)$, the missingness mechanism is
\emph{missing completely at random}(MCAR). In this case, complete-case analysis yields consistent 
estimates of the joint distribution $P(Y, M_1, M_2 \mid T, X)$.

When the missingness of each mediator depends only on the observed treatment, outcome, and 
covariates, the mechanism is \emph{missing at random}(MAR). Under MAR, the joint distribution 
$P(Y, M_1, M_2 \mid T, X)$ remains identifiable given the observed data.

In many practical situations, however, the missingness of a mediator $M_i$ may also depend on its 
own unobserved value, even after conditioning on the observed variables. Such a mechanism is 
termed MNAR, and it generally complicates identification. We consider four MNAR mechanisms of 
increasing complexity. In the first, the missingness of $M_i$ depends only on its own value and 
observed covariates, that is, $R_i$ depends on $(M_i, X)$. The second allows dependence on both 
$M_i$ and the treatment, $R_i$ depending on $(M_i, T, X)$. The third permits dependence on the 
outcome, so that $R_i$ depends on $(M_i, Y, X)$. The fourth and most general mechanism allows 
the missingness of $M_i$ to depend jointly on its own value, the treatment, and the outcome,i.e., 
$R_i$ depends on $(M_i, Y, T, X)$. Under these assumptions, the four mechanisms can be 
 formulated using either potential or observed outcomes respectively, and the corresponding 
identification results will be presented in the next section.

\section{Identifiability under MNAR mechanisms}
\begin{figure}[H]
\centering

\begin{subfigure}[t]{0.32\textwidth}
\centering
\begin{tikzpicture}[
    node distance=3mm and 18mm, 
    every node/.style={font=\small}
]
    \node (T) {$T$};
    \node (Y) [right=of T] {$Y$};
    \node (M1) [above=of T, xshift=11mm] {$M_1$};
    \node (R1) [above=of M1] {$R_1$};
    \node (M2) [below=of T, xshift=11mm] {$M_2$};
    \node (R2) [below=of M2] {$R_2$};

    \draw[->] (T) -- (Y);
    \draw[->] (T) -- (M1);
    \draw[->] (M1) -- (Y);
    \draw[->] (T) -- (M2);
    \draw[->] (M2) -- (Y);
\end{tikzpicture}
\caption*{(a)MCAR}
\end{subfigure}
\hfill
\begin{subfigure}[t]{0.32\textwidth}
\centering
\begin{tikzpicture}[
    node distance=3mm and 18mm, 
    every node/.style={font=\small} 
]
    \node (T) {$T$};
    \node (Y) [right=of T] {$Y$};
    \node (M1) [above=of T, xshift=11mm] {$M_1$};
    \node (R1) [above=of M1] {$R_1$};
    \node (M2) [below=of T, xshift=11mm] {$M_2$};
    \node (R2) [below=of M2] {$R_2$};

    \draw[->] (T) -- (Y);
    \draw[->] (T) -- (M1);
    \draw[->] (M1) -- (Y);
    \draw[->] (T) -- (M2);
    \draw[->] (M2) -- (Y);
  
    \draw[->] (T) -- (R1);
    \draw[->] (Y) -- (R1);
    \draw[->] (T) -- (R2);
    \draw[->] (Y) -- (R2);
\end{tikzpicture}
\caption*{(b)MAR}
\end{subfigure}
\hfill
\begin{subfigure}[t]{0.32\textwidth}
\centering
\begin{tikzpicture}[
    node distance=3mm and 18mm, 
    every node/.style={font=\small} 
]
    \node (T) {$T$};
    \node (Y) [right=of T] {$Y$};
    \node (M1) [above=of T, xshift=11mm] {$M_1$};
    \node (R1) [above=of M1] {$R_1$};
    \node (M2) [below=of T, xshift=11mm] {$M_2$};
    \node (R2) [below=of M2] {$R_2$};

    \draw[->] (T) -- (Y);
    \draw[->] (T) -- (M1);
    \draw[->] (M1) -- (Y);
    \draw[->] (T) -- (M2);
    \draw[->] (M2) -- (Y);
    \draw[->] (M1) -- (R1);
    \draw[->] (M2) -- (R2);
\end{tikzpicture}
\caption*{(c)MNAR mechanism 1}
\end{subfigure}

\vspace{6mm}

\begin{subfigure}[t]{0.32\textwidth}
\centering
\begin{tikzpicture}[
    node distance=3mm and 18mm, 
    every node/.style={font=\small} 
]
    \node (T) {$T$};
    \node (Y) [right=of T] {$Y$};
    \node (M1) [above=of T, xshift=11mm] {$M_1$};
    \node (R1) [above=of M1] {$R_1$};
    \node (M2) [below=of T, xshift=11mm] {$M_2$};
    \node (R2) [below=of M2] {$R_2$};

    \draw[->] (T) -- (Y);
    \draw[->] (T) -- (M1);
    \draw[->] (M1) -- (Y);
    \draw[->] (T) -- (M2);
    \draw[->] (M2) -- (Y);
    \draw[->] (T) -- (R1);
    \draw[->] (M1) -- (R1);
    \draw[->] (T) -- (R2);
    \draw[->] (M2) -- (R2);
\end{tikzpicture}
\caption*{(d)MNAR mechanism 2}
\end{subfigure}
\hfill
\begin{subfigure}[t]{0.32\textwidth}
\centering
\begin{tikzpicture}[
    node distance=3mm and 18mm,
    every node/.style={font=\small} 
]
    \node (T) {$T$};
    \node (Y) [right=of T] {$Y$};
    \node (M1) [above=of T, xshift=11mm] {$M_1$};
    \node (R1) [above=of M1] {$R_1$};
    \node (M2) [below=of T, xshift=11mm] {$M_2$};
    \node (R2) [below=of M2] {$R_2$};

    \draw[->] (T) -- (Y);
    \draw[->] (T) -- (M1);
    \draw[->] (M1) -- (Y);
    \draw[->] (T) -- (M2);
    \draw[->] (M2) -- (Y);
    \draw[->] (M1) -- (R1);
    \draw[->] (Y) -- (R1);
    \draw[->] (M2) -- (R2);
    \draw[->] (Y) -- (R2);
\end{tikzpicture}
\caption*{(e)MNAR mechanism 3}
\end{subfigure}
\hfill
\begin{subfigure}[t]{0.32\textwidth}
\centering
\begin{tikzpicture}[
    node distance=3mm and 18mm, 
    every node/.style={font=\small} 
]
    \node (T) {$T$};
    \node (Y) [right=of T] {$Y$};
    \node (M1) [above=of T, xshift=11mm] {$M_1$};
    \node (R1) [above=of M1] {$R_1$};
    \node (M2) [below=of T, xshift=11mm] {$M_2$};
    \node (R2) [below=of M2] {$R_2$};

    \draw[->] (T) -- (Y);
    \draw[->] (T) -- (M1);
    \draw[->] (M1) -- (Y);
    \draw[->] (T) -- (M2);
    \draw[->] (M2) -- (Y);
    \draw[->] (T) -- (R1);
    \draw[->] (M1) -- (R1);
    \draw[->] (Y) -- (R1);
    \draw[->] (T) -- (R2);
    \draw[->] (M2) -- (R2);
    \draw[->] (Y) -- (R2);
\end{tikzpicture}
\caption*{(f)MNAR mechanism 4}
\end{subfigure}
\caption{
Directed acyclic graphs (DAGs) illustrating the assumed missing mechanisms when missingness 
occurs only in the mediators. Panels (a)--(f) represent, in order, MCAR, MAR, and four 
increasingly general MNAR mechanisms. All graphs are conditioned on $X$, which may have directed 
edges to all variables.}
\label{fig:dag-missingness}
\end{figure}
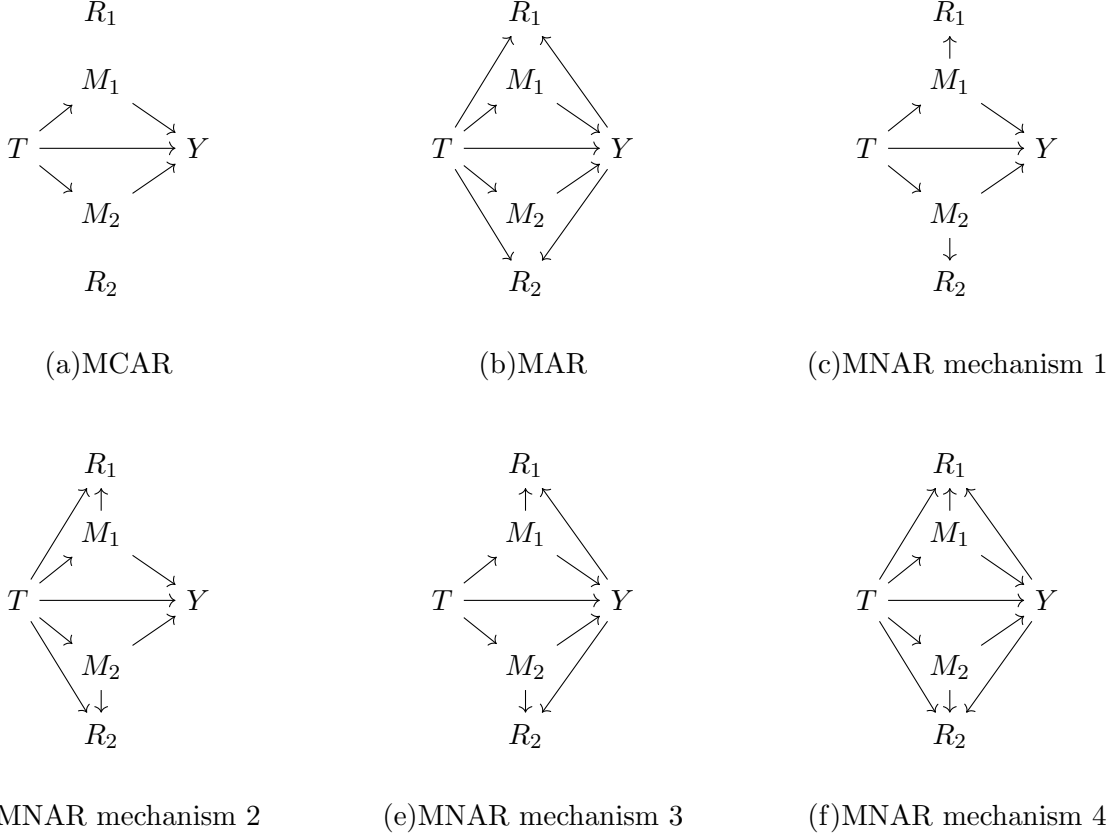

We consider four classes of MNAR mechanisms, corresponding to Mechanisms~1--4 in Figure~\ref{fig:dag-missingness}. 
The DAGs in Figure~\ref{fig:dag-missingness} provide schematic representations of the missingness mechanisms by highlighting the functional parents of the response indicators. 
Throughout, these graphs are interpreted conditional on $X$, whose arrows to the relevant variables are omitted for graphical clarity. 
The MNAR mechanisms differ in the conditioning sets that determine the response indicators $R_i$, that is, the variables on which the missingness of $M_i$ is allowed to depend. 
We further assume that each response indicator depends only on its own functional parents specified by the corresponding DAG and is conditionally independent of all non-parent variables given those parents. 
Consequently, the joint missingness mechanism factorizes as
$P(R_1,R_2 \mid M_1,M_2,Y,T,X) = P(R_1 \mid \operatorname{pa}(R_1))P(R_2 \mid \operatorname{pa}(R_2))$,
where $\operatorname{pa}(R_i)$ denotes the set of  parents of $R_i$.

\subsection{Identifiability under MNAR mechanism~1}

\begin{mechanism}
\label{mech:MNAR1}
$R_1 \perp\!\!\!\perp (T, Y, M_2, R_2) \mid (M_1, X)$, \quad
$R_2 \perp\!\!\!\perp (T, Y, M_1, R_1) \mid (M_2, X)$.
\end{mechanism}

Mechanism~1 assumes that the missingness indicator $R_i$ depends only on  
mediator $M_i$ and baseline covariates $X$. Conditional on $(M_i, X)$, $R_i$ is independent of
the treatment $T$, the outcome $Y$, and the other mediator $M_j$. This setting characterizes 
a basic MNAR scenario in which missingness is primarily driven by individual-level mediator 
values, rather than directly by the treatment or the outcome. In the context of the NHANES 
data, blood pressure and inflammatory markers are collected from different examination modules. 
Under this mechanism, whether an individual completes a given measurement may depend on their 
latent blood pressure or inflammatory status, for example, due to physical discomfort leading 
to non-participation. After conditioning on the corresponding mediator, the missingness 
processes across different modules can be reasonably approximated as conditionally independent.
Under these conditions, the following nonparametric identifiability result holds.

Suppose that $M_1$, $M_2$, and $Y$ take $I$, $J$, and $K$ distinct values, respectively, 
where $K\ge I$ and $K\ge J$ are required for identification, 
with supports $\mathscr M_1=\{m_{11},\ldots,m_{1I}\}$, $\mathscr M_2=\{m_{21},\ldots,m_{2J}\}$,
and $\mathscr Y=\{y_1,\ldots,y_K\}$. 

Define
\[
P^{(11)}_{m_{1i}m_{2j}y_k\mid t,x}
= P(R_1=1, R_2=1, M_1=m_{1i}, M_2=m_{2j}, Y=y_k \mid T=t,X=x),
\]
\[
\Theta_1(t,x)=\left(\sum_{m_2}P^{(11)}_{m_{1i}m_2y_k\mid t,x}\right)_{k,i}, \Theta_2(t,x)=\left(\sum_{m_1}P^{(11)}_{m_1m_{2j}y_k\mid t,x}\right)_{k,j}.
\]

\begin{theorem}
\label{thm:MNAR1}
Under Mechanism~1, 
the joint distribution $P(Y,M_1,M_2\mid T,X)$ is nonparametrically identifiable 
if the following conditions hold:
\begin{enumerate}
    \item[(1)] \textbf{Positivity conditions:} $P(R_1=1\mid M_1=m_1,X=x)>0$ and $P(R_2=1\mid M_2=m_2,X=x)>0$ for all $m_1\in\mathscr M_1$, $m_2\in\mathscr M_2$, and $x\in\mathscr X$;
    \item[(2)] \textbf{Rank conditions:} $\mathrm{rank}\{\Theta_1(t,x)\}=I$ and $\mathrm{rank}\{\Theta_2(t,x)\}=J$ for all $(t,x)$.
\end{enumerate}
Consequently, under Assumptions~1--4, the NDE and NIE are identifiable.
\end{theorem}

\begin{remark}
For continuously distributed mediators, analogous identification results may be established 
under suitable completeness conditions on the corresponding conditional distributions.
\end{remark}

\subsection{Identifiability under MNAR mechanism~2}

\begin{mechanism}
\label{mech:MNAR2}
$R_1 \perp\!\!\!\perp (Y, M_2, R_2) \mid (M_1, T, X)$, \quad
$R_2 \perp\!\!\!\perp (Y, M_1, R_1) \mid (M_2, T, X)$.
\end{mechanism}

Mechanism~2 extends Mechanism 1 by allowing the missingness mechanism to depend on the 
treatment $T$. Specifically, conditional on $(M_i, T, X)$, the missingness indicator $R_i$ is 
independent of the outcome $Y$ and the other mediator $M_j$. This corresponds to a 
treatment-driven MNAR setting, where missingness may be influenced not only by mediator values 
but also by treatment assignment. In practice, participation in certain examinations 
(e.g., blood tests) may depend not only on latent physiological conditions such as 
inflammation or blood pressure, but also on exposure status (e.g., obesity). For instance, 
severely obese individuals may exhibit lower compliance due to mobility limitations or health 
concerns.

\begin{theorem}
\label{thm:MNAR2}
Under Mechanism~2, 
the joint distribution $P(Y,M_1,M_2\mid T,X)$ is nonparametrically identifiable 
if the following conditions hold:
\begin{enumerate}
    \item[(1)] \textbf{Positivity conditions:} $P(R_1 = 1 \mid M_1 = m_1, T=t, X = x) > 0$ and $P(R_2 = 1 \mid M_2 = m_2, T=t, X = x) > 0$ for all $m_1\in\mathscr M_1$, $m_2\in\mathscr M_2$, $t\in\{0,1\}$, and $x\in\mathscr X$;
    \item[(2)] \textbf{Rank conditions:} $\mathrm{rank}\{\Theta_1(t,x)\}=I$ and $\mathrm{rank}\{\Theta_2(t,x)\}=J$ for all $(t,x)$, where necessarily $K\ge I$ and $K\ge J$.
\end{enumerate}
Consequently, under Assumptions~1--4, the NDE and NIE are identifiable.
\end{theorem}

\begin{remark}
Similar to Mechanism 1, for continuously distributed mediators, identification holds under 
suitable completeness conditions.
\end{remark}

\subsection{Identifiability under MNAR mechanism~3}

\begin{mechanism}
\label{mech:MNAR3}
$R_1 \perp\!\!\!\perp (T, M_2, R_2) \mid (M_1, X, Y)$, \quad
$R_2 \perp\!\!\!\perp (T, M_1, R_1) \mid (M_2, X, Y)$.
\end{mechanism}

Mechanism~3 further allows the missingness mechanism to depend on the outcome $Y$. Under this 
setting, $R_i$ may depend on $(M_i, Y, X)$, but is conditionally independent of the treatment 
$T$ and the other mediator $M_j$. This corresponds to an outcome-driven MNAR scenario. In NHANES, 
an individual’s health outcome (e.g., kidney function impairment) may affect their ability or 
willingness to participate in subsequent examinations. In addition, abnormal mediator levels may 
further increase the probability of missingness.

\begin{theorem}
\label{thm:MNAR3}
Suppose that $M_1$ and $M_2$ are binary.
Under Mechanism 3, 
the joint distribution $P(Y,M_1,M_2\mid T,X)$ is nonparametrically identifiable 
if the following conditions hold:
\begin{enumerate}
    \item[(1)] \textbf{Positivity conditions:} $P(R_1=1 \mid M_1=m_1, X=x, Y=y) > 0$ and $P(R_2=1 \mid M_2=m_2, X=x, Y=y) > 0$ for all $m_1\in\mathscr M_1$, $m_2\in\mathscr M_2$, $y\in\mathscr Y$, and $x\in\mathscr X$;
    \item[(2)] \textbf{Relevance conditions:} The treatment variable is conditionally associated with each mediator, that is, $T \not\!\perp\!\!\!\perp M_1 \mid (X, Y)$ and $T \not\!\perp\!\!\!\perp M_2 \mid (X, Y)$.
\end{enumerate}
Consequently, under Assumptions~1--4, the NDE and NIE are identifiable.
\end{theorem}

\subsection{Identifiability under MNAR mechanism~4}

\begin{mechanism}
\label{mech:MNAR4}
$R_1 \perp\!\!\!\perp (M_2, R_2) \mid (M_1, Y, T, X)$,\quad
$R_2 \perp\!\!\!\perp (M_1, R_1) \mid (M_2, Y, T, X)$.
\end{mechanism}

Mechanism~4 represents the most general setting considered in this paper. Here, $R_i$ may depend 
on $(M_i, T, Y, X)$, and is only assumed to be conditionally independent of the other mediator 
$M_j$. Compared with Mechanisms 1--3, this specification substantially relaxes exclusion 
restrictions and better reflects complex real-world missingness processes. However, under this 
general setting, the full data distribution is typically not nonparametrically identifiable from 
observed data alone. Additional structural assumptions are therefore required.

We impose the following parametric model on the missingness mechanism.

\begin{assumption}
\label{assump:logistic model}
The missingness mechanisms follow the following logistic regression models:
\[
\operatorname{logit} P(R_1 = 0 \mid T = t, M_1 = m_1, X = x, Y = y)
= \alpha_0 + \alpha_T t + \alpha_{M_1} m_1 + \alpha_X x + \alpha_Y y,
\]
\[
\operatorname{logit} P(R_2 = 0 \mid T = t, M_2 = m_2, X = x, Y = y)
= \beta_0 + \beta_T t + \beta_{M_2} m_2 + \beta_X x + \beta_Y y,
\]
where $\operatorname{logit}(a) = \log\{a/(1-a)\}$.
\end{assumption}

Under Assumption~\ref{assump:logistic model}, we can eliminate the parameters
$\alpha_0$, $\alpha_T$, $\alpha_X$, and $\alpha_Y$
in the first logistic model using the observed frequencies as estimated probabilities,
and then obtain two quadratic equations in $B=e^{\alpha_{M_1}}$:
\[
F_1B^2+G_1B+H_1=0, \quad F_2B^2+G_2B+H_2=0,
\]
where the coefficients $(F_i,G_i,H_i)$, $i=1,2$, are identifiable from the observed-data:
\[
F_1 = \frac{p_{1+0|10}^{(11)}p_{1+1|01}^{(11)}}{p_{++0|10}^{(01)}p_{++1|01}^{(01)}} 
- \frac{p_{1+1|00}^{(11)}p_{1+0|11}^{(11)}}{p_{++1|00}^{(01)}p_{++0|11}^{(01)}}
\]

\[
H_1 = \frac{p_{0+0|10}^{(11)}p_{0+1|01}^{(11)}}{p_{++0|10}^{(01)}p_{++1|01}^{(01)}} 
- \frac{p_{0+1|00}^{(11)}p_{0+0|11}^{(11)}}{p_{++1|00}^{(01)}p_{++0|11}^{(01)}}
\]

\[
G_1 = \frac{p_{0+0|10}^{(11)}p_{1+1|01}^{(11)}+p_{1+0|10}^{(11)}p_{0+1|01}^{(11)}}{p_{++0|10}^{(01)}p_{++1|01}^{(01)}}  
- \frac{p_{1+1|00}^{(11)}p_{0+0|11}^{(11)}+p_{0+1|00}^{(11)}p_{1+0|11}^{(11)}}{p_{++1|00}^{(01)}p_{++0|11}^{(01)}}
\]

\[
F_2 = \frac{p_{1+0|00}^{(11)}p_{1+1|11}^{(11)}}{p_{++0|00}^{(01)}p_{++1|11}^{(01)}} 
- \frac{p_{1+0|01}^{(11)}p_{1+1|10}^{(11)}}{p_{++0|01}^{(01)}p_{++1|10}^{(01)}}
\]

\[
H_2 = \frac{p_{0+0|00}^{(11)}p_{0+1|11}^{(11)}}{p_{++0|00}^{(01)}p_{++1|11}^{(01)}} 
- \frac{p_{0+0|01}^{(11)}p_{0+1|10}^{(11)}}{p_{++0|01}^{(01)}p_{++1|10}^{(01)}}
\]

\[
G_2 = \frac{p_{0+0|00}^{(11)}p_{1+1|11}^{(11)}+p_{1+0|00}^{(11)}p_{0+1|11}^{(11)}}{p_{++0|00}^{(01)}p_{++1|11}^{(01)}}
- \frac{p_{0+0|01}^{(11)}p_{1+1|10}^{(11)}+p_{1+0|01}^{(11)}p_{0+1|10}^{(11)}}{p_{++0|01}^{(01)}p_{++1|10}^{(01)}}
\]

Similarly, two quadratic equations are obtained for $B'=e^{\beta_{M_2}}$:
\[
F_1'B'^2+G_1'B'+H_1'=0, \quad F_2'B'^2+G_2'B'+H_2'=0.
\]
where the coefficients are identified as 
\[
F_1' = \frac{p_{+10|10}^{(11)}p_{+11|01}^{(11)}}{p_{++0|10}^{(10)}p_{++1|01}^{(10)}} 
- \frac{p_{+11|00}^{(11)}p_{+10|11}^{(11)}}{p_{++1|00}^{(10)}p_{++0|11}^{(10)}}
\]

\[
H_1' = \frac{p_{+00|10}^{(11)}p_{+01|01}^{(11)}}{p_{++0|10}^{(10)}p_{++1|01}^{(10)}} 
- \frac{p_{+01|00}^{(11)}p_{+00|11}^{(11)}}{p_{++1|00}^{(10)}p_{++0|11}^{(10)}}
\]

\[
G_1' = \frac{p_{+00|10}^{(11)}p_{+11|01}^{(11)}+p_{+10|10}^{(11)}p_{+01|01}^{(11)}}{p_{++0|10}^{(10)}p_{++1|01}^{(10)}}  
- \frac{p_{+11|00}^{(11)}p_{+00|11}^{(11)}+p_{+01|00}^{(11)}p_{+10|11}^{(11)}}{p_{++1|00}^{(10)}p_{++0|11}^{(10)}}
\]

\[
F_2' = \frac{p_{+10|00}^{(11)}p_{+11|11}^{(11)}}{p_{++0|00}^{(10)}p_{++1|11}^{(10)}} 
- \frac{p_{+10|01}^{(11)}p_{+11|10}^{(11)}}{p_{++0|01}^{(10)}p_{++1|10}^{(10)}}
\]

\[
H_2' = \frac{p_{+00|00}^{(11)}p_{+01|11}^{(11)}}{p_{++0|00}^{(10)}p_{++1|11}^{(10)}} 
- \frac{p_{+00|01}^{(11)}p_{+01|10}^{(11)}}{p_{++0|01}^{(10)}p_{++1|10}^{(10)}}
\]

\[
G_2' = \frac{p_{+00|00}^{(11)}p_{+11|11}^{(11)}+p_{+10|00}^{(11)}p_{+01|11}^{(11)}}{p_{++0|00}^{(10)}p_{++1|11}^{(10)}}
- \frac{p_{+00|01}^{(11)}p_{+11|10}^{(11)}+p_{+10|01}^{(11)}p_{+01|10}^{(11)}}{p_{++0|01}^{(10)}p_{++1|10}^{(10)}}
\]

Detailed derivations are provided in the Supplementary Materials.

\begin{theorem}
\label{thm:MNAR4}
Suppose that $M_1$, $M_2$, $Y$ and $X$ are all binary.
Under Mechanism~4 and Assumption~\ref{assump:logistic model},
the joint distribution $P(Y,M_1,M_2 \mid T,X)$ is identifiable
if the following conditions hold:
\begin{enumerate}
    \item[(1)] \textbf{Positivity conditions:} $P(R_1 = 1 \mid M_1 = m_1, T=t, X = x, Y=y) > 0$ and $P(R_2 = 1 \mid M_2 = m_2, T=t, X = x, Y=y) > 0$ for all $m_1\in\mathscr M_1$, $m_2\in\mathscr M_2$, $t\in\{0,1\}$, $x\in\mathscr X$ and $y\in\mathscr Y$;
    \item[(2)] \textbf{Uniqueness of the common positive root:}
    Let
    \[
    \mathcal S_B = \left\{B>0: F_1B^2+G_1B+H_1=0, \; F_2B^2+G_2B+H_2=0 \right\},
    \]
    then $\mathcal S_B$ contains exactly one element.
    This unique common positive root equals $B=e^{\alpha_{M_1}}$.

    Likewise, let
    \[
    \mathcal S_{B'} = \left\{B'>0: F_1'B'^2+G_1'B'+H_1'=0, \; F_2'B'^2+G_2'B'+H_2'=0 \right\},
    \]
    then $\mathcal S_{B'}$ contains exactly one element.
    This unique common positive root equals $B'=e^{\beta_{M_2}}$.
\end{enumerate}

Consequently, under Assumptions~1--4, the NDE and NIE are identifiable.
\end{theorem}

\begin{remark}
\label{remark:multiple_roots}
The uniqueness requirement in Condition~(2) is essential for point identification.

For the parameter $B=e^{\alpha_{M_1}}$, three situations may arise:

\begin{enumerate}
    \item[(i)]
    $\mathcal S_B=\varnothing$.
    In this case, the two quadratic equations have no common positive root.
    Hence no value of $B=e^{\alpha_{M_1}}$ is compatible with the observed-data,
    indicating either model misspecification or incompatibility between the assumed model
    and the observed-data distribution.

    \item[(ii)]
    $|\mathcal S_B|=1$.
    In this case, $B=e^{\alpha_{M_1}}$ is uniquely determined and is therefore point identified.

    \item[(iii)]
    $|\mathcal S_B|=2$.
    In this case, multiple positive values
    $B_1,B_2$
    satisfy both quadratic equations.
    Each solution corresponds to a distinct value of
    $\alpha_{M_1}=\log(B)$
    and generates the same observed-data distribution.
    Therefore $\alpha_{M_1}$ is not point identified and is only partially identifiable.
\end{enumerate}

An analogous classification applies to
$B'=e^{\beta_{M_2}}$ and the solution set
$\mathcal S_{B'}$.
\end{remark}

\section{Simulation}
We conducted a Monte Carlo simulation study to evaluate the finite-sample performance of the 
proposed method and to assess the validity of the theoretical results.

\subsection{Simulation Setup}
Let the baseline covariate $X$ and treatment $T$ be independently generated from a Bernoulli 
distribution with probability 0.5. The binary mediators $M_1$ and $M_2$, and the binary outcome 
$Y$, are generated according to the following logistic structural equations:
\begin{align*}
\text{logit} \, P(M_1=1 \mid T, X) &= \alpha_{10} + \alpha_{1t}T + \alpha_{1x}X, \\
\text{logit} \, P(M_2=1 \mid T, X) &= \alpha_{20} + \alpha_{2t}T + \alpha_{2x}X, \\
\text{logit} \, P(Y=1 \mid T, M_1, M_2, X) &= \beta_0 + \beta_t T + \beta_1 M_1 + \beta_2 M_2 + \beta_{12} M_1 M_2 + \beta_x X.
\end{align*}

We consider four missingness mechanisms corresponding to Mechanisms~1--4 described in Section~4.
For each mechanism, the response indicator $R_i$ depends on the variables specified by the 
corresponding DAG and conditional independence assumptions. Specifically, we generate the binary 
variable $R_i$ from
\[
\text{logit} \, P(R_i=1 \mid M_i,X) = \gamma_{i0} + \gamma_{i1}M_i + \gamma_{ix}X;
\]
under mechanism 1;
\[
\text{logit} \, P(R_i=1 \mid M_i,T,X) = \gamma_{i0} + \gamma_{i1}M_i + \gamma_{it}T + \gamma_{ix}X;
\]
under mechanism 2;
\[
\text{logit} \, P(R_i=1 \mid M_i,T,Y) = \gamma_{i0} + \gamma_{i1}M_i + \gamma_{it}T + \gamma_{iy}Y;
\]
under mechanism 3;
\[
\text{logit} \, P(R_i=1 \mid M_i,T,Y,X) = \gamma_{i0} + \gamma_{i1}M_i + \gamma_{it}T + \gamma_{iy}Y + \gamma_{ix}X.
\]
under mechanism 4.

The parameters $\gamma$ are calibrated such that the missing rates of the mediators are 
approximately 20\%--30\%, comparable to those observed in NHANES. For each scenario, we generate 
independent and identically distributed samples of size $n=2000$, and repeat the simulation 500 
times.

\subsection{Simulation Results}
We compare the following methods for estimating the NIE, NDE, and path-specific effects:
(1)Complete-case analysis (CC);
(2)Multiple imputation by chained equations (MICE) under the MAR assumption;
(3)The proposed method in this paper(MNAR\_EM);
(4)Oracle estimators, which are obtained by using the true values of the missing data.

Performance is evaluated in terms of bias, standard deviation (SD), and root mean squared error 
(RMSE). Figure 2 presents the percentage bias of each method across the four missingness
mechanisms.

\begin{figure}[htbp]
    \centering

    \begin{subfigure}[b]{0.48\textwidth}
        \centering
        \includegraphics[width=\textwidth]{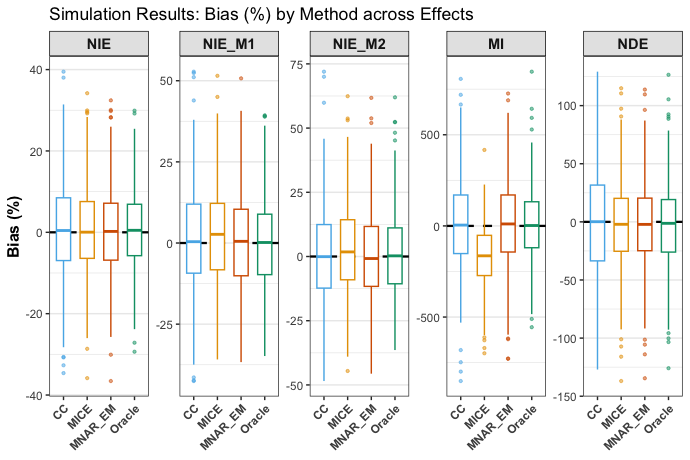} 
        \caption{MNAR mechanism 1}
        \label{fig:mech1}
    \end{subfigure}
    \hfill
    \begin{subfigure}[b]{0.48\textwidth}
        \centering
        \includegraphics[width=\textwidth]{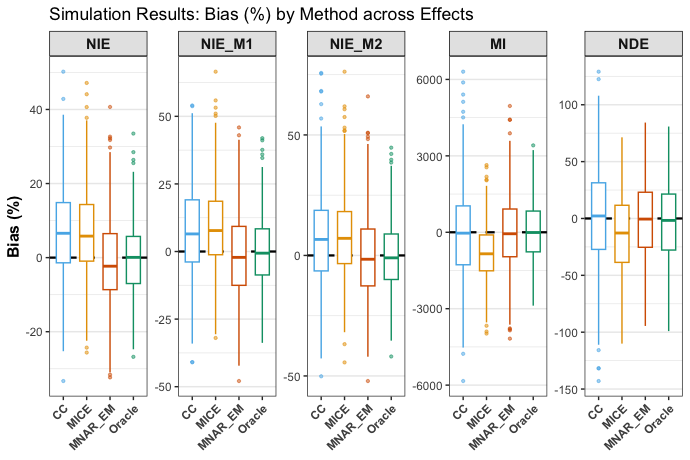}
        \caption{MNAR mechanism 2}
        \label{fig:mech2}
    \end{subfigure}

    \vspace{6mm}

    \begin{subfigure}[b]{0.48\textwidth}
        \centering
        \includegraphics[width=\textwidth]{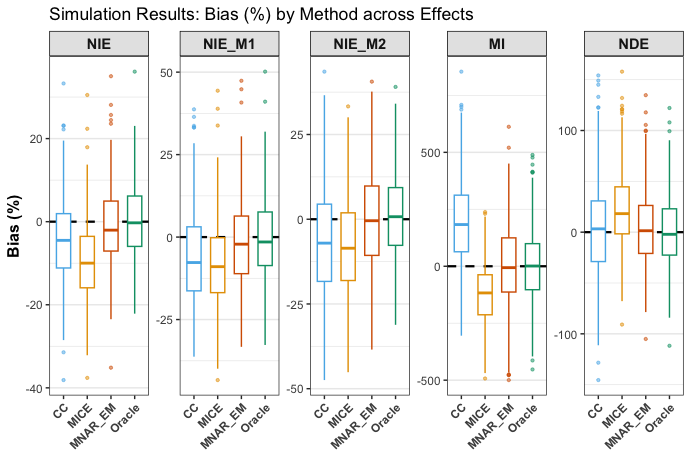}
        \caption{MNAR mechanism 3}
        \label{fig:mech3}
    \end{subfigure}
    \hfill
    \begin{subfigure}[b]{0.48\textwidth}
        \centering
        \includegraphics[width=\textwidth]{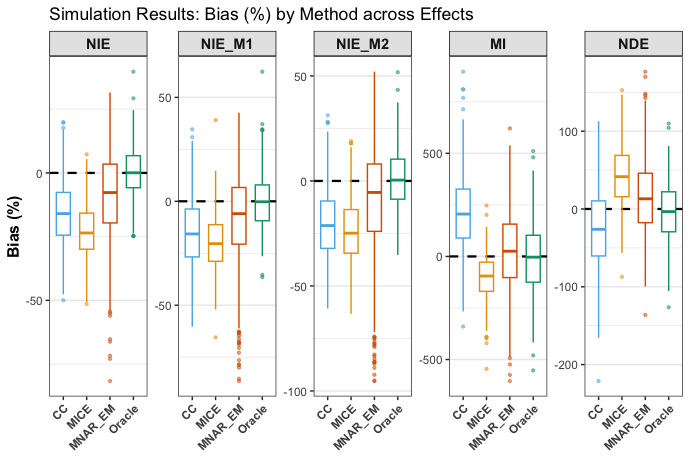}
        \caption{MNAR mechanism 4}
        \label{fig:mech4}
    \end{subfigure}

    \caption{
        Boxplots of percentage bias for the estimated effects across four methods (CC, MICE, 
        MNAR-EM, and Oracle) under four missingness mechanisms. Results are based on 500 Monte 
        Carlo replications with $n=2000$.
    }
    \label{fig:boxplot_mechanisms}
\end{figure}

Overall, the Oracle estimator exhibited negligible bias across all scenarios, confirming the 
validity of the simulation design. Under relatively simple mechanisms (Mechanisms 1 and 2), 
CC and MICE showed small but noticeable systematic bias, indicating that ignoring the missingness 
mechanism may still introduce non-negligible errors even under mild MNAR settings. In contrast, 
the proposed method in this paper consistently achieved small bias and produced estimates close 
to the Oracle benchmark. Its variability was comparable to that of MICE and slightly lower than 
that of CC, resulting in superior overall performance in terms of RMSE.

As the complexity of the missingness mechanism increased (Mechanisms 3 and 4), the performance 
gap between methods widened substantially. Under Mechanism 3, where missingness depends on the 
outcome, the key assumptions underlying CC and MICE are violated, leading to increased bias and 
RMSE. The MNAR-EM method, however, maintained relatively low bias and stable performance. Under 
the most complex setting (Mechanism 4), where missingness depends jointly on multiple variables, 
CC and MICE exhibited substantial bias and instability across effect estimates. Although the bias 
of MNAR-EM increased to some extent, it remained markedly smaller than that of the competing 
methods.

In summary, when the missingness mechanism deviates from MAR-particularly when it depends on the 
outcome or multiple variables—standard methods such as CC and MICE may suffer from substantial 
bias. The MNAR-EM we propose, by explicitly modeling the missingness process, effectively 
reduces bias across a range of MNAR scenarios and achieves performance close to the Oracle 
benchmark.

\section{Application}

\subsection{Data}
To demonstrate the practical utility of the proposed method, we investigate the causal effect of 
obesity on renal function impairment and its underlying pathways using data from the National 
Health and Nutrition Examination Survey (NHANES). Specifically, this section focuses on the core 
scientific question: whether obesity increases the risk of renal function impairment, and to what 
extent this effect is transmitted through two parallel pathways—hypertension and systemic 
inflammation. While NHANES is a nationally representative cross-sectional survey, its physical 
examination and laboratory testing components are characterized by voluntary participation, which 
introduces a potential source of non-randomness in missing mediator values.

The analysis sample consists of 2,443 adults, for whom the treatment variable, outcome variable, 
and baseline covariates are fully observed. The binary treatment variable $T$ is an indicator for 
obesity (BMI $\geq 30\,\text{kg/m}^2$ is recorded as 1). The outcome variable $Y$ represents renal 
function impairment status, defined as 1 when the urinary albumin-to-creatinine ratio (ACR) 
$\geq 30\,\text{mg/g}$. The mediator $M_1$ is a hypertension indicator: it takes the value 1 if 
the average of the second and third measurements of systolic blood pressure $\geq 130\,\text{mmHg}$ 
or diastolic blood pressure $\geq 80\,\text{mmHg}$, and 0 otherwise. The mediator $M_2$ is a 
systemic inflammation indicator: it takes the value 1 if the white blood cell count exceeds 
$10 \times 10^3\,\text{cells}/\mu\text{L}$, and 0 otherwise. The vector of baseline covariates 
$\bm{X}$ includes age, sex, and the family income-to-poverty ratio.

The data show a certain proportion of missingness in both mediators: the missing rate for 
hypertension $M_1$ is 14.74\%, and for inflammation $M_2$ is 5.49\%; samples with at least one 
missing mediator account for 18.71\%. In NHANES, the completion of physiological measurements is 
often directly influenced by the participants' own health status; for instance, individuals with 
poorer health or mobility issues caused by severe obesity typically show lower compliance. Such a 
missing mechanism may depend directly on the latent true values of the variables, thereby 
deviating from the MAR assumption and exhibiting MNAR characteristics. Therefore, methods based 
on complete case analysis or the MAR assumption may lead to systemic bias.

In addition, to assess the empirical plausibility of treating hypertension and systemic 
inflammation as causally non-ordered parallel mediators, we further examined whether the two 
mediators remained conditionally associated after adjusting for treatment and covariates. We 
fitted two logistic regression models among complete cases, $P(M_2 \mid M_1,T,X)$ and 
$P(M_1 \mid M_2,T,X)$. The results showed that the association between hypertension and 
systemic inflammation was not statistically significant in either direction. Specifically, in 
the model for $P(M_2 \mid M_1,T,X)$, the odds ratio was 1.122 (95\% CI: 0.755--1.668, $p=0.569$); 
in the model for $P(M_1 \mid M_2,T,X)$, the odds ratio was 1.154 (95\% CI: 0.779--1.709, 
$p=0.477$). These results indicate that, among complete cases, there was no clear empirical 
evidence of conditional association between the two mediators after conditioning on treatment 
and covariates, thereby supporting the plausibility of treating them as causally non-ordered 
parallel mediators in the empirical analysis.

\subsection{Model Selection}
Based on the four types of MNAR mechanisms proposed in Section 3, we employ the EM algorithm for 
maximum likelihood estimation. Utilizing the parametric nested relationships among the missingness 
models, we compare different mechanisms via the likelihood ratio test (LRT) for model selection. 
Table~\ref{tab:model_selection} reports the number of parameters, log-likelihood values at 
convergence, information criteria, and LRT results for each model.

\begin{table}[htbp]
  \centering
  \caption{Model comparison and likelihood ratio tests for four parametric MNAR mechanisms}
  \label{tab:model_selection}
  \begin{tabular}{lccccc}
    \hline
    Mechanism & $k$ & Log-Likelihood & AIC & LRT (df) & $p$-value \\
    \hline
    Mechanism 1
      & 28 & $-4021.506$ & 8099.013  & -- & -- \\
    Mechanism 2
      & 30 & $-4016.448$ & 8092.896 
      & 10.117 (2)\textsuperscript{a} & 0.0064 \\
    Mechanism 3
      & 30 & $-4021.040$ & 8102.080 
      & 0.932 (2)\textsuperscript{b} & 0.6274 \\
    Mechanism 4
      & 32 & $-4015.953$ & 8095.905 
      & 0.991 (2)\textsuperscript{c} & 0.6094 \\
    \hline
    \multicolumn{6}{l}{\footnotesize Note: LRT is based on nested model comparisons.} \\
    \multicolumn{6}{l}{\footnotesize \textsuperscript{a} Mechanism 2 vs Mechanism 1;} \\
    \multicolumn{6}{l}{\footnotesize \textsuperscript{b} Mechanism 3 vs Mechanism 1;} \\
    \multicolumn{6}{l}{\footnotesize \textsuperscript{c} Mechanism 4 vs Mechanism 2.}
  \end{tabular}
\end{table}

The LRT results in Table 1 indicate that: (i) introducing the treatment variable $T$ (Mechanism 2) based 
on Mechanism 1 significantly improves model fit ($p=0.006$); (ii) introducing the outcome variable 
$Y$ (Mechanism 3) based on Mechanism 1 does not yield significant improvement ($p=0.627$); (iii) 
further introducing $Y$ (Mechanism 4) based on Mechanism 2 also fails to significantly improve the 
fit ($p=0.609$). Therefore, we consider Mechanism 2 to be most consistent with the data, which 
also aligns with prior epidemiological knowledge. In health surveys, a participant's willingness 
to undergo physical examinations is often directly influenced by their known obesity status and 
associated physiological and psychological concerns; they are more likely to refrain from 
examinations due to fears of revealing more serious health conditions, rather than depending 
directly on their unobserved renal function status. Based on this statistical evidence and 
scientific plausibility, we adopt Mechanism 2 as the primary analysis model.

\subsection{Results}
Under Mechanism 2, we estimate the causal mediation effects using CC, MICE, and MNAR\_EM, 
respectively, and construct confidence intervals via non-parametric Bootstrap ($B = 500$). 
Table~\ref{tab:main_results} reports the estimation results for each method.

\begin{table}[htbp]
\centering
\caption{Comparison of causal mediation effect estimates under different methods (Bootstrap $B=500$)}
\label{tab:main_results}
\begin{tabular}{lcccccc}
\hline
& \multicolumn{2}{c}{CC}
& \multicolumn{2}{c}{MICE}
& \multicolumn{2}{c}{MNAR\_EM} \\
\cline{2-3} \cline{4-5} \cline{6-7}
Effect & Estimate & 95\% CI
     & Estimate & 95\% CI
     & Estimate & 95\% CI \\
\hline
NIE
& 0.0088 & [0.0028, 0.0155]
& 0.0080 & [0.0026, 0.0146]
& 0.0080 & [0.0034, 0.0134] \\

NDE
& -0.0097 & [-0.0392, 0.0227]
& -0.0056 & [-0.0340, 0.0178]
& -0.0060 & [-0.0279, 0.0153] \\

NIE$_{M_1}$
& 0.0052 & [0.0017, 0.0099]
& 0.0044 & [0.0013, 0.0095]
& 0.0045 & [0.0017, 0.0078] \\

NIE$_{M_2}$
& 0.0024 & [-0.0026, 0.0080]
& 0.0025 & [-0.0019, 0.0079]
& 0.0024 & [-0.0011, 0.0064] \\

MI
& 0.0013 & [0.0002, 0.0027]
& 0.0011 & [0.0001, 0.0026]
& 0.0011 & [0.0002, 0.0021] \\
\hline
\end{tabular}
\end{table}

Under the proposed method, the NIE is $0.0080$ (95\% CI: $[0.0034, 0.0134]$), indicating that 
obesity increases the risk of renal damage by approximately 0.80 percentage points through the 
combined mediating roles of hypertension and inflammation. The estimate for the NDE is $-0.0060$ 
(95\% CI: $[-0.0279, 0.0153]$), where the confidence interval includes zero, suggesting that 
there is no sufficient evidence to support a direct causal effect of obesity on renal damage 
under this mechanism.

Further decomposition of the mediation pathways reveals that: the indirect effect through 
hypertension ($\text{NIE}_{M_1} = 0.0045$; 95\% CI: $[0.0017, 0.0078]$) is statistically 
significant, suggesting that hypertension may be a crucial mediator through which obesity affects 
renal damage. The indirect effect through inflammation ($\text{NIE}_{M_2} = 0.0024$; 
95\% CI: $[-0.0011, 0.0064]$), though not reaching statistical significance, has a positive point 
estimate consistent with existing biological pathological mechanisms. Additionally, the mediated 
interaction effect (MI = $0.0011$; 95\% CI: $0.0002$ to $0.0021$) is positive and significant, 
indicating a potential synergistic effect between the two pathways; that is, the risk of renal 
damage when both hypertension and inflammation are present is slightly higher than the simple 
sum of their independent effects.

Comparing the results across methods, the point estimates obtained from the three methods are 
generally close, indicating that the main conclusions are relatively robust in this application. 
However, the complete case analysis and multiple imputation methods rely on MCAR or MAR 
assumptions, respectively, which lack sufficient empirical support in the current application 
and may introduce potential bias. In contrast, the proposed method explicitly models the MNAR 
mechanism, grounding the inference in data-generating assumptions that are closer to reality. 
While the point estimates are largely consistent, the confidence intervals obtained by the 
proposed method are generally more concentrated, suggesting that after appropriately specifying 
the missingness mechanism, the observed information can be utilized more fully, thereby improving 
estimation precision to a certain extent.

\section{Discussion}

This paper studies the identifiability of causal mediation effects with multiple parallel 
mediators under MNAR mechanisms. By embedding non-ignorable missingness into the 
potential-outcomes framework for mediation analysis, we derive identification results under 
four progressively generalized MNAR structures. Specifically, under the weaker MNAR mechanisms 
(Mechanisms 1--3), nonparametric identification is achieved under completeness and support 
conditions; under the most general Mechanism 4, identification requires parametric specification 
of the missingness mechanism. These results extend existing mediation analysis to settings 
where mediator missingness depends on latent variables.

Despite their theoretical generality, the identification results rely on structural assumptions 
that are not empirically testable. In particular, Assumptions 1--4 impose sequential ignorability 
and cross-world independence, which are fundamental to the definition of natural direct and 
indirect effects but cannot be verified from observed data. Consequently, the validity of the 
proposed identification results depends on domain knowledge rather than statistical testability.

Under Mechanisms 1--3, nonparametric identification hinges on completeness and support conditions, 
which ensure invertibility of conditional expectation operators. However, such conditions are 
typically difficult to verify in practice and may fail when the support of the mediators is 
limited or weakly informative, leading to a gap between theoretical identification and empirical 
applicability. For Mechanism 4, identification relies on a logistic specification of the 
missingness mechanism, making the results inherently model-dependent: misspecification of the 
missingness model may induce bias even if the causal structure is correctly specified.

In addition, the framework assumes independence among multiple mediators at the level of potential 
outcomes, which facilitates identification of parallel mediation paths. It should be noted, 
however, that the decomposition of natural indirect effects already implicitly imposes a 
path-separability structure. This provides a natural basis for extension to sequential mediation 
settings, where mediators follow a temporal or causal ordering. In such cases, interaction terms 
admit clearer path-specific interpretations. Therefore, the decomposition results for MI can be 
viewed as a building block for ordered mediation models.

Finally, the choice among different MNAR mechanisms is not identifiable from observed data alone. 
Although model selection in the empirical analysis is guided by likelihood-based criteria and 
domain knowledge, alternative MNAR specifications cannot be ruled out. This highlights the 
inherent model uncertainty under MNAR settings. Future work should develop sensitivity analysis 
frameworks to characterize the robustness of mediation effect estimates over a class of 
admissible missingness mechanisms, rather than relying on a single selected model.

\section*{Supplementary Materials}
Proofs are provided in the Supplementary Materials.
\par

\section*{Acknowledgements}
The authors acknowledge all forms of support that contributed to the completion of this research. 
The authors are grateful to Professor Peng Ding from the Department of Statistics, University of 
California, Berkeley, for his valuable comments and constructive suggestions.
\par

\section*{Disclosure statement}
No potential conflict of interest was reported by the author(s).
\par

\section*{Funding}
This work was supported by the BTBU Digital Business Platform Project funded by BMEC at Beijing 
Technology and Business University.
\par

\bibhang=1.7pc
\bibsep=2pt
\fontsize{9}{14pt plus.8pt minus .6pt}\selectfont
\renewcommand\bibname{\large \bf References}
\expandafter\ifx\csname
natexlab\endcsname\relax\def\natexlab#1{#1}\fi
\expandafter\ifx\csname url\endcsname\relax
  \def\url#1{\texttt{#1}}\fi
\expandafter\ifx\csname urlprefix\endcsname\relax\def\urlprefix{URL}\fi

\bibliographystyle{chicago}      
\bibliography{references}   

\begin{thebibliography}{}

\bibitem[\protect\citeauthoryear{Campion}{Campion}{1989}]{Rubin2004}
Campion, W.~M. (1989).
\newblock {\em Book review: multiple imputation for nonresponse in surveys}.
\newblock SAGE Publications Sage CA: Los Angeles, CA.

\bibitem[\protect\citeauthoryear{Chen and Lin}{Chen and Lin}{2025}]{Chen2025}
Chen, Y.-L. and S.-H. Lin (2025).
\newblock Definition and interpretation of separable path-specific effects with multiple ordered mediators.
\newblock {\em Epidemiology\/}~{\em 36\/}(5), 677--685.

\bibitem[\protect\citeauthoryear{Daniel, De~Stavola, Cousens, and Vansteelandt}{Daniel et~al.}{2015}]{Daniel2015}
Daniel, R.~M., B.~L. De~Stavola, S.~N. Cousens, and S.~Vansteelandt (2015).
\newblock Causal mediation analysis with multiple mediators.
\newblock {\em Biometrics\/}~{\em 71\/}(1), 1--14.

\bibitem[\protect\citeauthoryear{Dashti, Lee, Simpson, Carlin, and Moreno-Betancur}{Dashti et~al.}{2025}]{Dashti2025}
Dashti, S.~G., K.~J. Lee, J.~A. Simpson, J.~B. Carlin, and M.~Moreno-Betancur (2025).
\newblock Handling multivariable missing data in causal mediation analysis estimating interventional effects.
\newblock {\em Epidemiology\/}~{\em 36\/}(4), 487--499.

\bibitem[\protect\citeauthoryear{Davidian}{Davidian}{2022}]{Davidian2022}
Davidian, M. (2022).
\newblock Methods based on semiparametric theory for analysis in the presence of missing data.
\newblock {\em Annual Review of Statistics and Its Application\/}~{\em 9}, 167--196.

\bibitem[\protect\citeauthoryear{Dempster, Laird, and Rubin}{Dempster et~al.}{1977}]{Dempster1977}
Dempster, A.~P., N.~M. Laird, and D.~B. Rubin (1977).
\newblock Maximum likelihood from incomplete data via the em algorithm.
\newblock {\em Journal of the royal statistical society: series B (methodological)\/}~{\em 39\/}(1), 1--22.

\bibitem[\protect\citeauthoryear{Ding and Geng}{Ding and Geng}{2014}]{Ding2014}
Ding, P. and Z.~Geng (2014).
\newblock Identifiability of subgroup causal effects in randomized experiments with nonignorable missing covariates.
\newblock {\em Statistics in Medicine\/}~{\em 33\/}(7), 1121--1133.

\bibitem[\protect\citeauthoryear{Huber}{Huber}{2014}]{Huber2014}
Huber, M. (2014).
\newblock Identifying causal mechanisms (primarily) based on inverse probability weighting.
\newblock {\em Journal of Applied Econometrics\/}~{\em 29\/}(6), 920--943.

\bibitem[\protect\citeauthoryear{Huber and Solovyeva}{Huber and Solovyeva}{2020}]{Huber2020}
Huber, M. and A.~Solovyeva (2020).
\newblock Direct and indirect effects under sample selection and outcome attrition.
\newblock {\em Econometrics\/}~{\em 8\/}(4), 44.

\bibitem[\protect\citeauthoryear{Imai, Keele, and Tingley}{Imai et~al.}{2010}]{Imai2010a}
Imai, K., L.~Keele, and D.~Tingley (2010).
\newblock A general approach to causal mediation analysis.
\newblock {\em Psychological methods\/}~{\em 15\/}(4), 309.

\bibitem[\protect\citeauthoryear{Imai, Keele, and Yamamoto}{Imai et~al.}{2010}]{Imai2010b}
Imai, K., L.~Keele, and T.~Yamamoto (2010).
\newblock Identification, inference and sensitivity analysis for causal mediation effects.
\newblock {\em Statistical Science\/}~{\em 25\/}(1), 51--71.

\bibitem[\protect\citeauthoryear{Kim}{Kim}{2011}]{Kim2011}
Kim, J.~K. (2011).
\newblock Parametric fractional imputation for missing data analysis.
\newblock {\em Biometrika\/}~{\em 98\/}(1), 119--132.

\bibitem[\protect\citeauthoryear{Li and Zhou}{Li and Zhou}{2017}]{Li2017}
Li, W. and X.-H. Zhou (2017).
\newblock Identifiability and estimation of causal mediation effects with missing data.
\newblock {\em Statistics in Medicine\/}~{\em 36\/}(25), 3948--3965.

\bibitem[\protect\citeauthoryear{Li, Miao, Shpitser, and Tchetgen~Tchetgen}{Li et~al.}{2023}]{Li2023}
Li, Y., W.~Miao, I.~Shpitser, and E.~J. Tchetgen~Tchetgen (2023).
\newblock A self-censoring model for multivariate nonignorable nonmonotone missing data.
\newblock {\em Biometrics\/}~{\em 79\/}(4), 3203--3214.

\bibitem[\protect\citeauthoryear{Little and Rubin}{Little and Rubin}{2019}]{Little2019}
Little, R.~J. and D.~B. Rubin (2019).
\newblock {\em Statistical analysis with missing data}.
\newblock John Wiley \& Sons.

\bibitem[\protect\citeauthoryear{Ma, Geng, and Hu}{Ma et~al.}{2003}]{Ma2003}
Ma, W.-Q., Z.~Geng, and Y.-H. Hu (2003).
\newblock Identification of graphical models for nonignorable nonresponse of binary outcomes in longitudinal studies.
\newblock {\em Journal of multivariate analysis\/}~{\em 87\/}(1), 24--45.

\bibitem[\protect\citeauthoryear{Mhasawade and Chunara}{Mhasawade and Chunara}{2024}]{Mhasawade2024}
Mhasawade, V. and R.~Chunara (2024).
\newblock Disparate effect of missing mediators on transportability of causal effects.
\newblock {\em arXiv preprint arXiv:2403.08638\/}.

\bibitem[\protect\citeauthoryear{Miao, Ding, and Geng}{Miao et~al.}{2016}]{Miao2016a}
Miao, W., P.~Ding, and Z.~Geng (2016).
\newblock Identifiability of normal and normal mixture models with nonignorable missing data.
\newblock {\em Journal of the American Statistical Association\/}~{\em 111\/}(516), 1673--1683.

\bibitem[\protect\citeauthoryear{Miao and Tchetgen~Tchetgen}{Miao and Tchetgen~Tchetgen}{2016}]{Miao2016b}
Miao, W. and E.~J. Tchetgen~Tchetgen (2016).
\newblock On varieties of doubly robust estimators under missingness not at random with a shadow variable.
\newblock {\em Biometrika\/}~{\em 103\/}(2), 475--482.

\bibitem[\protect\citeauthoryear{Mohan and Pearl}{Mohan and Pearl}{2021}]{Mohan2021}
Mohan, K. and J.~Pearl (2021).
\newblock Graphical models for processing missing data.
\newblock {\em Journal of the American Statistical Association\/}~{\em 116\/}(534), 1023--1037.

\bibitem[\protect\citeauthoryear{Moreno, Wu, Yap, Lam, Wetter, Nahum-Shani, Dempsey, and Rehg}{Moreno et~al.}{2020}]{Moreno2020}
Moreno, A., Z.~Wu, J.~R. Yap, C.~Lam, D.~Wetter, I.~Nahum-Shani, W.~Dempsey, and J.~M. Rehg (2020).
\newblock A robust functional em algorithm for incomplete panel count data.
\newblock {\em Advances in neural information processing systems\/}~{\em 33}, 19828--19838.

\bibitem[\protect\citeauthoryear{Pearl}{Pearl}{2022}]{Pearl2001}
Pearl, J. (2022).
\newblock Direct and indirect effects.
\newblock In {\em Probabilistic and causal inference: the works of Judea Pearl}, pp.\  373--392.

\bibitem[\protect\citeauthoryear{Preacher and Hayes}{Preacher and Hayes}{2004}]{Preacher2004}
Preacher, K.~J. and A.~F. Hayes (2004).
\newblock Spss and sas procedures for estimating indirect effects in simple mediation models.
\newblock {\em Behavior research methods, instruments, \& computers\/}~{\em 36\/}(4), 717--731.

\bibitem[\protect\citeauthoryear{Robins, Rotnitzky, and Zhao}{Robins et~al.}{1994}]{Robins1994}
Robins, J.~M., A.~Rotnitzky, and L.~P. Zhao (1994).
\newblock Estimation of regression coefficients when some regressors are not always observed.
\newblock {\em Journal of the American statistical Association\/}~{\em 89\/}(427), 846--866.

\bibitem[\protect\citeauthoryear{Robins, Rotnitzky, and Zhao}{Robins et~al.}{1995}]{Robins1995}
Robins, J.~M., A.~Rotnitzky, and L.~P. Zhao (1995).
\newblock Analysis of semiparametric regression models for repeated outcomes in the presence of missing data.
\newblock {\em Journal of the american statistical association\/}~{\em 90\/}(429), 106--121.

\bibitem[\protect\citeauthoryear{Rubin}{Rubin}{1976}]{Rubin1976}
Rubin, D.~B. (1976).
\newblock Inference and missing data.
\newblock {\em Biometrika\/}~{\em 63\/}(3), 581--592.

\bibitem[\protect\citeauthoryear{Seaman and White}{Seaman and White}{2013}]{Seaman2013}
Seaman, S.~R. and I.~R. White (2013).
\newblock Review of inverse probability weighting for dealing with missing data.
\newblock {\em Statistical methods in medical research\/}~{\em 22\/}(3), 278--295.

\bibitem[\protect\citeauthoryear{Sengewald, Hardt, and Sengewald}{Sengewald et~al.}{2025}]{Sengewald2024}
Sengewald, E., K.~Hardt, and M.-A. Sengewald (2025).
\newblock A causal view on bias in missing data imputation: The impact of evil auxiliary variables on norming of test scores.
\newblock {\em Multivariate Behavioral Research\/}~{\em 60\/}(2), 258--274.

\bibitem[\protect\citeauthoryear{Shan, Li, and Ai}{Shan et~al.}{2026}]{Shan2026}
Shan, J., W.~Li, and C.~Ai (2026).
\newblock Efficient nonparametric inference for mediation analysis with nonignorable missing confounders.
\newblock {\em Journal of the American Statistical Association\/}~(just-accepted), 1--22.

\bibitem[\protect\citeauthoryear{Sun, Liu, Miao, Wirth, Robins, and Tchetgen}{Sun et~al.}{2018}]{Sun2018a}
Sun, B., L.~Liu, W.~Miao, K.~Wirth, J.~Robins, and E.~J.~T. Tchetgen (2018).
\newblock Semiparametric estimation with data missing not at random using an instrumental variable.
\newblock {\em Statistica Sinica\/}~{\em 28\/}(4), 1965.

\bibitem[\protect\citeauthoryear{Sun, Perkins, Cole, Harel, Mitchell, Schisterman, and Tchetgen~Tchetgen}{Sun et~al.}{2018}]{Sun2018b}
Sun, B., N.~J. Perkins, S.~R. Cole, O.~Harel, E.~M. Mitchell, E.~F. Schisterman, and E.~J. Tchetgen~Tchetgen (2018).
\newblock Inverse-probability-weighted estimation for monotone and nonmonotone missing data.
\newblock {\em American journal of epidemiology\/}~{\em 187\/}(3), 585--591.

\bibitem[\protect\citeauthoryear{Taguri, Featherstone, and Cheng}{Taguri et~al.}{2018}]{Taguri2018}
Taguri, M., J.~Featherstone, and J.~Cheng (2018).
\newblock Causal mediation analysis with multiple causally non-ordered mediators.
\newblock {\em Statistical methods in medical research\/}~{\em 27\/}(1), 3--19.

\bibitem[\protect\citeauthoryear{Tchetgen and Shpitser}{Tchetgen and Shpitser}{2012}]{Tchetgen2012}
Tchetgen, E. J.~T. and I.~Shpitser (2012).
\newblock Semiparametric theory for causal mediation analysis: efficiency bounds, multiple robustness, and sensitivity analysis.
\newblock {\em Annals of statistics\/}~{\em 40\/}(3), 1816.

\bibitem[\protect\citeauthoryear{Tsiatis}{Tsiatis}{2006}]{Tsiatis2006}
Tsiatis, A.~A. (2006).
\newblock {\em Semiparametric theory and missing data}.
\newblock Springer.

\bibitem[\protect\citeauthoryear{Valeri and VanderWeele}{Valeri and VanderWeele}{2013}]{Valeri2013}
Valeri, L. and T.~J. VanderWeele (2013).
\newblock Mediation analysis allowing for exposure--mediator interactions and causal interpretation: theoretical assumptions and implementation with sas and spss macros.
\newblock {\em Psychological methods\/}~{\em 18\/}(2), 137.

\bibitem[\protect\citeauthoryear{van Buuren}{van Buuren}{2018}]{vanBuuren2018}
van Buuren, S. (2018).
\newblock {\em Flexible Imputation of Missing Data, Second Edition}.
\newblock Chapman \& Hall/CRC.

\bibitem[\protect\citeauthoryear{VanderWeele and Vansteelandt}{VanderWeele and Vansteelandt}{2014}]{VanderWeele2014}
VanderWeele, T. and S.~Vansteelandt (2014).
\newblock Mediation analysis with multiple mediators.
\newblock {\em Epidemiologic methods\/}~{\em 2\/}(1), 95--115.

\bibitem[\protect\citeauthoryear{Vansteelandt, Bekaert, and Lange}{Vansteelandt et~al.}{2012}]{Vansteelandt2012b}
Vansteelandt, S., M.~Bekaert, and T.~Lange (2012).
\newblock Imputation strategies for the estimation of natural direct and indirect effects.
\newblock {\em Epidemiologic Methods\/}~{\em 1\/}(1), 131--158.

\bibitem[\protect\citeauthoryear{Wang, Shao, and Kim}{Wang et~al.}{2014}]{Wang2014}
Wang, S., J.~Shao, and J.~K. Kim (2014).
\newblock An instrumental variable approach for identification and estimation with nonignorable nonresponse.
\newblock {\em Statistica Sinica\/}, 1097--1116.

\bibitem[\protect\citeauthoryear{Xia and Chan}{Xia and Chan}{2022}]{Xia2022}
Xia, F. and K.~C.~G. Chan (2022).
\newblock Decomposition, identification and multiply robust estimation of natural mediation effects with multiple mediators.
\newblock {\em Biometrika\/}~{\em 109\/}(4), 1085--1100.

\bibitem[\protect\citeauthoryear{Yang, Lorch, and Small}{Yang et~al.}{2014}]{Yang2014a}
Yang, F., S.~A. Lorch, and D.~S. Small (2014).
\newblock Estimation of causal effects using instrumental variables with nonignorable missing covariates: application to effect of type of delivery nicu on premature infants.
\newblock {\em The Annals of Applied Statistics\/}, 48--73.

\bibitem[\protect\citeauthoryear{Yang, Wang, and Ding}{Yang et~al.}{2019}]{Yang2019}
Yang, S., L.~Wang, and P.~Ding (2019).
\newblock Causal inference with confounders missing not at random.
\newblock {\em Biometrika\/}~{\em 106\/}(4), 875--888.

\bibitem[\protect\citeauthoryear{Zhou}{Zhou}{2022}]{Zhou2022}
Zhou, X. (2022).
\newblock Semiparametric estimation for causal mediation analysis with multiple causally ordered mediators.
\newblock {\em Journal of the Royal Statistical Society Series B: Statistical Methodology\/}~{\em 84\/}(3), 794--821.

\bibitem[\protect\citeauthoryear{Zuo, Ghosh, Ding, and Yang}{Zuo et~al.}{2025}]{Zuo2025}
Zuo, S., D.~Ghosh, P.~Ding, and F.~Yang (2025).
\newblock Mediation analysis with the mediator and outcome missing not at random.
\newblock {\em Journal of the American Statistical Association\/}~{\em 120\/}(550), 794--804.

\end{thebibliography}

\vskip .65cm
\noindent
 authors affiliation \\
School of Mathematics and Statistics, Beijing Technology and Business University
\vskip 2pt
\noindent
E-mail: (2431051078@st.btbu.edu.cn)\\
E-mail: (shanshanluo@btbu.edu.cn)\\
E-mail: (xlwang@btbu.edu.cn)\\
E-mail: (zhigeng@pku.edu.cn)
\vskip 2pt

\end{document}